# Ultrafast photodetection in the quantum wells of single AlGaAs/GaAs-based nanowires

N. Erhard[1,2], S. Zenger[1], S. Morkötter[1], D. Rudolph[1], M. Weiss[2,3], H. J. Krenner[2,3], H. Karl[3], G. Abstreiter[1,2,4], J. J. Finley[1,2], G. Koblmüller[1], and A. W. Holleitner[*,1,2]

[1] *Walter Schottky Institut and Physics Department, Technical University of Munich, D-85748 Garching, Germany*

[2] *Nanosystems Initiative Munich (NIM), Schellingstr. 4, D-80799 München, Germany*

[3] *Institute of Physics, Universität Augsburg, Universitätsstr. 1, D-86135 Augsburg, Germany*

[4] *Institute for Advanced Study, Technische Universität München, D-85748 Garching, Germany*

**Abstract:**

We investigate the ultrafast optoelectronic properties of single $Al_{0.3}Ga_{0.7}As$/GaAs-core-shell-nanowires. The nanowires contain GaAs-based quantum wells. For a resonant excitation of the quantum wells, we find a picosecond photocurrent which is consistent with an ultrafast lateral expansion of the photogenerated charge carriers. This Dember-effect does not occur for an excitation of the GaAs-based core of the nanowires. Instead, the core exhibits an ultrafast displacement current and a photo-thermoelectric current at the metal Schottky contacts. Our results uncover the optoelectronic dynamics in semiconductor core-shell nanowires comprising quantum wells, and they demonstrate the possibility to use the low-dimensional quantum well states therein for ultrafast photoswitches and photodetectors.

Semiconducting nanowires have attracted increasing attention as building blocks for optoelectronic devices such as solar cells,[1][2][3] photodetectors,[4][5][6][7] LEDs,[8][9][10] single photon emitters,[11][12][13] and lasers.[14][15] Nanowires further allow to integrate quantum confined electron states into semiconductor circuits. Such states can have two-dimensional (2D)[16][17], one-dimensional (1D),[16][17][18] and zero-dimensional (0D) character[11][12][13][18][19][20][22], which are isolated from the free surface and host the active device region. Among the various different material systems, GaAs-AlGaAs nanowires are a particularly interesting model system for radial 2D quantum wells (QWs), since the embedded QWs are nearly lattice-matched to the matrix material with negligible strain and piezoelectric effects.[23] So far, GaAs-AlGaAs based QWs in nanowires have been studied with respect to their optical, electronic and structural properties using e.g. photoluminescence spectroscopy[3] and inelastic light scattering.[2] However, investigation of the carrier dynamics and electrical transport properties are largely hampered for NWs consisting of both a bulk-like (3D) GaAs core and a (2D) QW. For all-electrical experiments, charge carriers have to be introduced by doping. Since multiple channels cannot be doped selectively, disentangling the transport characteristic of multiple channels is highly non-trivial.[24] Here, we apply a picosecond photocurrent spectroscopy to single GaAs-AlGaAs based QW-nanowires.[25] Since the core and the QWs are made out of GaAs in such nanowires, an optical excitation generates electrons and holes in both. We reveal that photocurrent spectroscopy allows to resolve the 3D joint density of states of the core states. In turn, the photocurrent response of the QWs shows up as a distinct, extra photocurrent response. Therefore, the QW states can be optoelectronically addressed and exploited for photodetection. Using this technique, we demonstrate that the QW states in GaAs-AlGaAs core-shell nanowires can be exploited in ultrafast photoswitches and photodetectors. In particular, we find that the quantum well states exhibit a picosecond optoelectronic response at cryogenic temperatures, which we interpret in terms of a lateral photo-Dember effect.[26][27] This effect originates from the different diffusivity of electrons and holes in III-V semiconductor components. The diffusivities are enhanced at low dimensions.[28] Therefore, the QW states are particularly suited for such a picosecond photocurrent response. We characterize the optoelectronic properties of the QW nanowires using energy-, space-, and time-resolved photocurrent measurements. Our results may pave the way to use quantum well states in core-shell nanowires for applications in high-speed optoelectronic circuits.

The 10 µm long nanowires are grown on a Si (111) substrate by molecular beam epitaxy (Figure 1a and supporting information).[17] Figure 1b schematically shows the QW-nanowire structure. The nominally undoped nanowires consist of a GaAs-core (green) with a diameter of 60-80 nm which is surrounded by a 30 nm thick $Al_{0.3}Ga_{0.7}As$-shell (light yellow), followed by a 2 nm thick GaAs-QW layer (green) and a 30 nm thick $Al_{0.3}Ga_{0.7}As$-shell (light yellow). A capping layer of 10 nm GaAs (green) prevents the oxidation of the outer AlGaAs-shell. The nanowires are mechanically transferred onto a preprocessed sapphire substrate and contacted by evaporated Ti/Au contacts with a thickness of 10 nm/90 nm without annealing (Figure 1c). The contacts form lateral, coplanar strip lines with a total length of ~58 mm, a width of 5 µm and a separation of 6 µm. The strip lines are needed to perform the time-resolved photocurrent experiments,[29][30][31] which are discussed below. Due to the nanowire composition, two different metal/nanowire contacts are formed: (i) At the nanowire base (white triangle in Figure 1c), the GaAs-core and the QWs are in direct contact to the Ti/Au. (ii) At the nanowire tip (black triangle), only the GaAs cap is in direct contact to the Ti/Au, while the GaAs-based QWs and core are further separated from the metals by the AlGaAs shell(s). The I-V-characteristics of the nanowire-circuits have an s-shape at room temperature (Figure 2a), which is consistent with having Schottky-contacts at both the nanowire base and the tip.[32] The overall dark current $I_{Dark}$ is small (~0.5 nA at 5 V), which can be explained by the fact that the nanowires are nominally undoped in combination with the presence of Schottky contacts. The dark current drops with decreasing temperature such that we do not measure any current signal above noise level for temperatures below 100 K.

In a next step, we detect two-dimensional photocurrent maps of the contacted NWs (Figure 2b and supporting information). In particular, we measure the time-averaged photocurrent $I_{Photo}$ at a laser energy of 1.7 eV, zero bias, and a bath temperature $T_{bath}$ of 8 K (Figure 2c); i.e. a temperature and bias at which no dark current is detectable. The chosen laser energy is smaller than the band gap of the AlGaAs-shells (~1.9 eV). Hereby, charge carriers are predominantly excited in the GaAs-core and the GaAs-QWs of the nanowires. We observe a positive time-averaged photocurrent close to the nanowire base (white triangle in Figure 2c). This finding is consistent with the expectation that at this position, the core and the QWs are in contact to the metals. At the nanowire tip (black triangle), the signal of the time-averaged photocurrent is not detectable by time-averaged photocurrent measurements. The details of the photocurrent generation mechanisms at both the base and the tip of the nanowire will be discussed below when we present the time-resolved photocurrent data. We note that the spatial dependence of the photocurrent in Figure 2c corresponds to the laser spot shape with a diameter of about 5 µm. From the photocurrent direction (photogenerated electrons moving to the left), we can deduce that the Schottky contact at the nanowire base is slightly *n*-type (supporting information).

To explore the dimensionality of the electronic states within the GaAs-core and the GaAs-QWs, we measure the time-averaged photocurrent $I_{Photo}$ as a function of the laser excitation energy (Figure 3a). For each energy, a spatially resolved photocurrent map is detected similar to the one in Figure 2c. The maps are fitted by two-dimensional Gaussians, and the corresponding maximum photocurrent is plotted as a function of the laser energy in Figure 3a. The depicted errors are fit errors. For energies below 1.5 eV, we do not observe any photocurrent signal which is consistent with the fact that GaAs has a band gap of ~1.51 eV at 10 K.[33] In the range between 1.5 eV and 1.7 eV, the photocurrent increases with excitation energy, exhibiting a square root dependence. This finding is consistent with a 3D joint density of states and with the fact that the GaAs-core diameter of 60-80 nm exceeds the Bohr radius of ~12 nm in GaAs. In other words, the GaAs-core exhibits 3D electronic states. At an excitation energy of ~1.68 eV, we observe a step-like increase of the photocurrent (red area in Figure 3a). We interpret it to correspond to the optical excitation of the QW-states. When we subtract the fitted square root dependence representing the GaAs-core from the overall photocurrent of the nanowire, we can extract an extra photocurrent within the quantum well $I_{QW}$ (inset of Figure 3a). Within the experimental error, $I_{QW}$ exhibits an $E^{-1/2}$ behavior which is characteristic for a 1D density of states. For a two-dimensional QW, one would expect a constant density of states. In fact, the GaAs-QW forms a hexagonal tube within the nanowire. Such a QW-tube exhibits electron states which are usually a superposition of 2D states (electrons confined to the six facets of the hexagonal QW-tube) and 1D states (electrons confined to the six edge-channels of the hexagonal QW tube). For the ground state, the electrons are confined to the 1D edge-channels of the QW-tube which principally matches the observed behavior of $I_{QW}$.[16][17] However, for thin QWs as present in our sample, thickness fluctuations and the associated localization of excitons in "natural" quantum dots are expected to play the dominant role.[18][34] For excitation energies just below the lowest QW transition energy, the absorption in the QW is enhanced due to the formation of quantum dot excitons. This effect is more pronounced in 2D structures like the GaAs-QWs and does not occur for the 3D GaAs core. Therefore, the energy dependence of $I_{QW}$ can also be explained by an enhanced absorption due to exciton effects in the QW. As shown in experiments performed on NWs from the same growth run presented in Ref.(34), carrier transport is well preserved in the QW continuum. Activation of carriers from interface fluctuations to the 2D continuum is highly efficient due to the shallow confinement of these QDs even at low temperatures. In a photocurrent experiment, absorption in such localized states occurs and gives rise to the observed deviation of the ideal 2D density of states. Thus we can conclude, that the measured photocurrent indeed comprises photogenerated, but dissociated electron-hole pairs in the QW. These charge carriers are in fact mobile, such that a current can be detected in the overall macroscopic circuit. Figure 3b shows the photoluminescence of another QW-nanowire from the same wafer. Indeed, the onset of the photoluminescence at 1.51 eV corresponds to the band gap of the GaAs-core. A set of emission peaks (red) occurs at ~1.7 eV which matches the lowest transition energy for the extra photocurrent from the QW (red area in Figure 3a). The sharp photoluminescence emission lines are consistent with the expectation of localized excitons. Localization may originate from a combination of local thickness and alloy fluctuations in the QW and crystal structure,[18][34][35] i.e. with twinning defects between wurtzite and zinc-blende segments. [36]

We now turn to the ultrafast optoelectronic dynamics occurring in the GaAs-core of the nanowires. We use an on-chip THz-time domain photocurrent spectroscopy, where an optical femtosecond pump laser excites the electronic states within the nanowire. This laser is the same as in the presented time-integrated photocurrent measurements in Figures 2c and 3a. Since the contacts form strip lines, the photocurrents give also rise to electromagnetic transients in the metal strip lines with a bandwidth of up to 2 THz. The transients run along the strip lines, and they are detected on-chip by a time-delayed optical femtosecond probe pulse in combination with an Auston-switch.[25] We use ion-implanted amorphous silicon for this ultrafast photodetector with a time-resolution of about 1 ps. [37][29][30][31] The current $I_{Sampling}$ across the Auston-switch is sampled as a function of the time-delay $\Delta t$ between the two laser pulses, and it is directly proportional to the ultrafast photocurrents in the nanowire.[29][30][31] Hereby, the photocurrents in the nanowire can be measured with a picosecond time-resolution. Figures 4a and 4b show $I_{Sampling}$ vs $\Delta t$ for resonant excitation of the GaAs-core (Fig. 4a with $E_{laser}$ = 1.53 eV at $T_{bath}$ = 8 K and Fig. 4b with $E_{laser}$ = 1.42 eV at $T_{bath}$ = 296 K). In each panel, the bottom trace corresponds to a laser excitation at the nanowire base (white triangle). The top trace relates to the nanowire tip (black triangle). The traces in-between correspond to the positions marked by black dots in Figure 2c. Each curve of $I_{Sampling}$ is fitted by a red line which has two independent fit-components: (i) A Lorentzian with a full width of half maximum (FWHM) of ~1.5 ps and (ii) an exponentially modified Gaussian with a decay time of ~5 ps. First, we discuss the Lorentzian response. It can be explained by an ultrafast displacement current with the following current density [29][30]

$$j_{\text{Displacement}} = \varepsilon_0 \varepsilon \frac{\partial E}{\partial t}, \tag{1}$$

with $\varepsilon$ ($\varepsilon_0$) the relative (vacuum) permittivity. In this ultrafast process, photogenerated charge carriers redistribute within the semiconductor nanowire to locally decrease electric fields $E$.[29][38][39] The data in Figure 4 are measured at zero bias. Since the sign of the fast displacement current is positive along the whole nanowire, we can deduce that an intrinsic electric field exists along the nanowire. This finding is consistent with a dominant Schottky barrier at the nanowire base as discussed in the context of Figure 2c. For nominally undoped nanowires, Schottky fields stretch far into the nanowires. The amplitude of the ultrafast displacement is given by the strength of the electric field in combination with the optical absorption at each position. The absorption is maximum in-between the two metal contacts, which explains the spatial amplitude variation of the ultrafast displacement current in Figures 4a and 4b. According to the fits (red lines), the second ultrafast photocurrent contribution can be described by an exponentially modified Gaussian. Intriguingly, the sign of this contribution does not change equally when the exciting laser is scanned across the nanowire (Figures 4a and 4b). This contribution is consistent with a photo-thermoelectric current generated at the nanowire base[29][31][40]

$$I_{\text{Thermo}} = (S_{\text{Nanowire}} - S_{\text{Stripline}}) \Delta T / R, \tag{2}$$

with $\Delta T$ the local temperature increase, $R$ the total resistance of the electrical circuit and $S_{\text{Nanowire}}$ ($S_{\text{Stripline}}$) the Seebeck coefficient of the nanowire (Ti/Au contact). At the nanowire base, a direct interface exists to the electron bath in the metal contact. Consistent with this interpretation, all presented data (time-resolved and time-integrated) are independent of the applied bias voltage in an investigated range of -5 V to +5 V. In particular, a photo-thermoelectric current $I_{\text{Thermo}}$ is not expected to have a bias dependence as the Seebeck-coefficients are independent of the bias voltage.

Intriguingly, the ultrafast photocurrent dynamics change at low temperatures, when the electronic states in the QWs are excited. Figure 4c depicts the time-resolved $I_{\text{Sampling}}$ for $E_{\text{laser}}$ = 1.59 eV at $T_{\text{Bath}}$ = 8 K, and Figure 4d for $E_{\text{laser}}$ = 1.70 eV and at $T_{\text{Bath}}$ = 296 K. At room-temperature, we observe qualitatively the same time-resolved photocurrent behavior as for the resonant excitation in the GaAs-core. In turn, the room-temperature photocurrent in the GaAs-QWs is consistent with an ultrafast displacement current and a photo-thermoelectric current generated at the nanowire base. At 8 K (Figure 4c), however, the photocurrent in the QW becomes negative for a laser excitation at the nanowire base (white triangle) and positive for an excitation at the nanowire tip (black triangle) (independent of the applied bias). The spatial sign change of the photocurrent is only observed at low temperatures. The origin of the optoelectronic dynamics in the QWs is revealed by power-dependent photocurrent measurements discussed in the next section.

We perform a power-dependence of the pump laser pulse at $T_{Bath}$ = 8 K and numerically compute the absolute area $A_{Sampling}$ below the time-resolved photocurrent $I_{Sampling}$ as in Figure 4. This amplitude is plotted in Figure 5a. When only the GaAs-core is excited ($E_{laser}$ = 1.53 eV), $A_{Sampling}$ exhibits an initial fast increase with a saturation for the pump laser power exceeding 100 µW. This saturation can be understood by an ultrafast displacement current which screens the electric field of a Schottky barrier.[41] The higher the laser power, the more charge carriers are generated to screen the electric field. In turn, the displacement current increases as a function of the laser power. As soon as sufficient charge carriers are photogenerated to totally screen the electric field in the NW, the displacement current saturates. We note that we detect an equivalent saturation behavior for all measurements at room temperature. Intriguingly, for an excitation of the QWs at low temperatures, we do not observe a saturation as a function of laser power. Instead, the amplitude of the photocurrent increases linearly with laser power in the examined range. The spatial sign change and the linear increase of the photocurrent in the QWs can be understood by a lateral photo-Dember effect.[27] Figures 5b and 5c schematically depict the photocurrent generation mechanism of such a photo-Dember effect. The exciting laser exhibits a Gaussian-like intensity distribution (red areas). The laser is partially shadowed by the Ti/Au contacts which are on top of the nanowire. Proportional to the incident laser intensity profile, charge carriers are photogenerated in the nanowire QW. Typically, the diffusivity of the electrons greatly exceeds the diffusivity of the holes in III-V semiconductor quantum wells.[28] Thereby, the electron cloud expands faster than the hole cloud (Figures 5b and 5c), which results in asymmetric electron and hole clouds shortly after the laser excitation. In turn, a so-called Dember current density $j_{Dember}$ is generated towards the contacts with opposite sign at the nanowire base and tip (arrows in Figures 5b and c). Hereby, one can explain the sign change of the ultrafast photocurrent for the nanowire being excited at its base or tip. In addition, the photo-Dember effect also explains that at the nanowire base (white triangle), $j_{Dember}$ has the opposite direction and sign than the displacement current. The latter has the direction of the

*n*-type Schottky field. In this interpretation, the photo-Dember effect has a maximum amplitude at the contacts, where the asymmetry of the electron and hole clouds is maximum due to the shadowing effect, although the optical absorption would be maximum in the center between the contacts. Consistently, at the center, we detect a superposition of ultrafast currents with cancelling directions (Figure 4c). We further note that a photo-Dember effect should give a linear power dependence when detected in a THz-time domain spectroscopy,[30] as we observe in Figure 5a.

Generally, the well-known Einstein relation is modified in degenerate electron systems.[42] In 1D and 2D systems, the diffusivity-mobility-ratio is larger than in bulk material and it increases with charge carrier density.[28] For illumination with a 500 µW and 160 fs laser pulse, the nanowire QW can be considered to be highly degenerate leading to a high diffusivity-mobility-ratio. Furthermore, the diffusivity in GaAs-QWs increases with photon energy,[43] and the diffusion length increases with decreasing temperature and increasing laser power.[44] Hence, diffusion effects such as the lateral photo-Dember effect are enhanced for QW excitation at low temperatures. In agreement with this interpretation, we observe the lateral photo-Dember effect in the QW only at 8 K but not at room-temperature. At room-temperature, the photogenerated electrons and holes are very likely emitted through a thermionic emission from the QW into the GaAs-core within ~1 ps (cf. supporting information). In turn, directly after the photoexcitation, all photocurrents generated in the QW are suppressed because of a thermionic transmission from the QW to the core. At 8 K, however, the thermionic emission rate is strongly reduced, and the photogenerated charge carrier dynamics predominantly occur within the QWs. We note that the nanowires are designed such that the tunneling rates through the 30 nm thick AlGaAs barrier are too low to significantly influence the charge carrier distribution in the QW (cf. supporting information). The thermionic emission explains why we only observe the lateral photo-Dember effect at 8 K.

In summary, we study the ultrafast photocurrents in AlGaAs/GaAs-core-shell nanowires containing 2 nm thick QWs. The photogenerated charge carriers enable a lateral photo-Dember effect at 8 K for laser excitation of the QWs. At room temperature, the data are consistent with photogenerated charge carriers being emitted within 1 ps to the GaAs-core. For a resonant excitation of the 3D core of the nanowire, other optoelectronic dynamics emerge. These are an ultrafast displacement current and a photo-thermoelectric current. Both currents originate at the interface of the nanowires to the metal contacts of the overall circuit.

The first current is driven by the electric fields at a Schottky barrier. The second one comprises the diffusion and drift of hot electrons to and from the interface. Our results reveal the optoelectronic dynamics in AlGaAs/GaAs-core-shell nanowires for ultrafast photoswitches and photodetectors.

**AUTHOR INFORMATION**

Corresponding Authors

*E-Mail: holleitner@wsi.tum.de

The authors declare no competing financial interest.

**ACKNOWLEDGEMENTS**

We gratefully acknowledge financial support of the ERC-grant NanoREAL, the Deutsche Forschungsgemeinschaft via the SFB 631, the Emmy Noether Program (H.J.K. and M.W., KR3790/2-1), the Center for NanoScience (CeNS), and the International Graduate School of Science and Engineering (IGSSE) of the Technische Universität München.

**Figure Captions:**

**Figure 1** (a) Scanning electron microscope (SEM) image and (b) schematic drawing of the QW-nanowires grown on a Si (111) substrate. The GaAs-core, the 2 nm QW and the GaAs capping layer are depicted in green. The AlGaAs inner and outer shell are shown in light yellow and the Si substrate in dark blue. (c) Schematic profile of the contacted QW-nanowire lying on a sapphire substrate. The hatched areas are the metal contacts to the nanowire. The black and white triangles mark the two different contact types at the top and the base of the nanowires, as discussed in the text.

**Figure 2** (a) Dark current voltage characteristic of a QW-nanowire contacted by two Ti/Au contacts ($T \sim$ 296 K). (b) Optical microscope image of an investigated QW-nanowire. (c) Time integrated photocurrent $I_{Photo}$ map of the nanowire. The dashed lines mark the position of the strip line contacts. Black dots indicate the nanowire orientation and the excitation positions for time-resolved photocurrent measurements ($T \sim$ 8 K, $U_{SD}$ = 0 V, $P_{Laser}$ = 500 µW, $E_{Laser}$ = 1.7 eV).

**Figure 3** (a) Time-integrated photocurrent $I_{Photo}$ as a function of excitation energy. The black line is a square root fit through the data points. The red area marks the photocurrent generated in the quantum wells, which is depicted in the inset as photocurrent $I_{QW}$. The red line in the inset has the shape of a 1D density of state function ($T \sim$ 8 K, $U_{SD}$ = 0 V, $P_{Laser}$ = 500 µW). (b) Photoluminescence signal of a second QW-nanowire grown on the same wafer ($T \sim$ 4.2 K, $E_{Laser}$ = 1.88 eV).

**Figure 4** Time-resolved photocurrent $I_{sampling}$ at $\sim$ 8 K for a laser excitation energy of (a) 1.53 eV and (b) 1.70 eV and at $\sim$ 296 K for (c) 1.42 eV and (d) 1.59 eV. The different traces correspond to the laser excitation positions along the nanowire (black dots in Figure 2) starting at the position marked by an open black triangle (bottom trace). The data are made offset for a better visibility ($U_{SD}$ = 0 V, $P_{Laser}$ = 500 µW, $P_{Probe,1.53eV}$ = 116 mW, $P_{Probe,1.70eV}$ = 70 mW, $P_{Probe,1.42eV}$ = 96 mW, $P_{Probe,1.59eV}$ = 96 mW).

**Figure 5** (a) Absolute integrated amplitude $A_{Sampling}$ of the time-resolved photocurrent $I_{sampling}$ as a function of laser power for an excitation at the position marked by a white triangle. The red line is a linear fit from the data obtained for excitation at 1.70 eV, the black line is a saturation fit from the data obtained for excitation at 1.53 eV. (b) and (c): Schematic drawing visualizing the lateral photo-Dember effect with the Dember current density $j_{Dember}$ for the position marked by black and white triangles, respectively.

**Figures:**

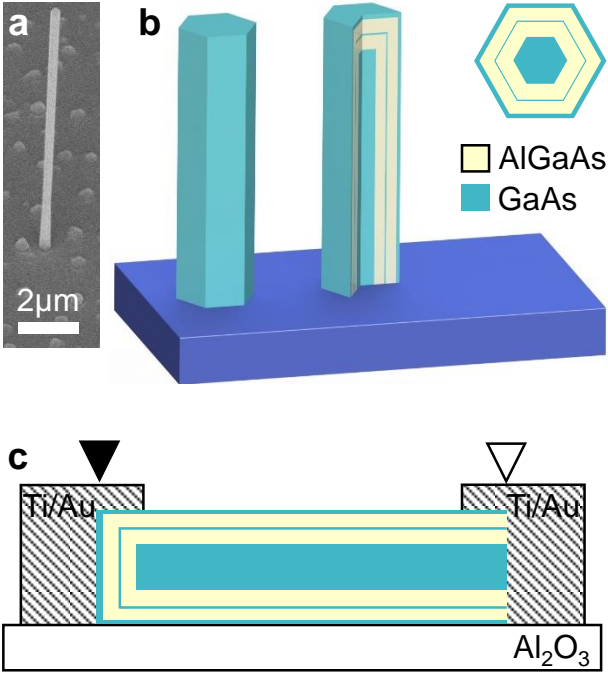

*Figure 1*

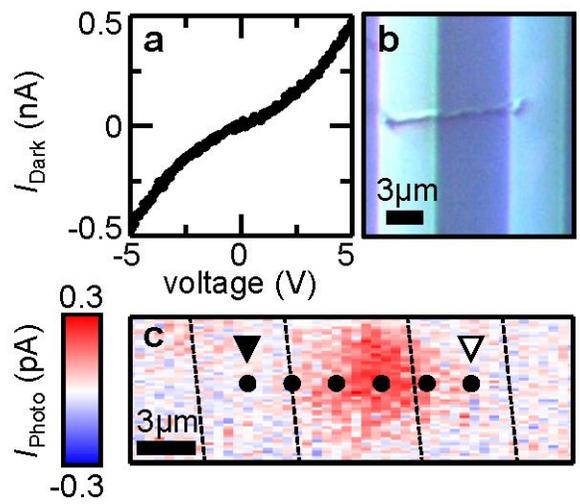

*Figure 2*

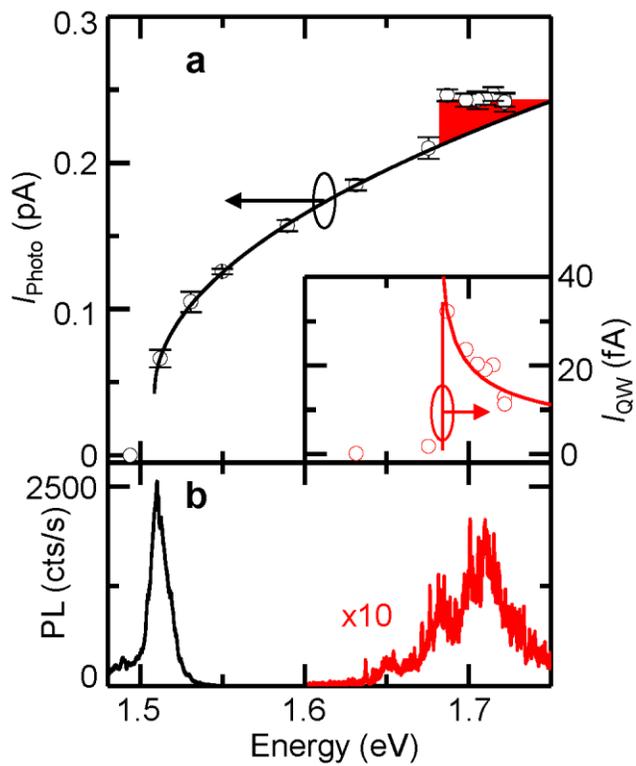

*Figure 3*

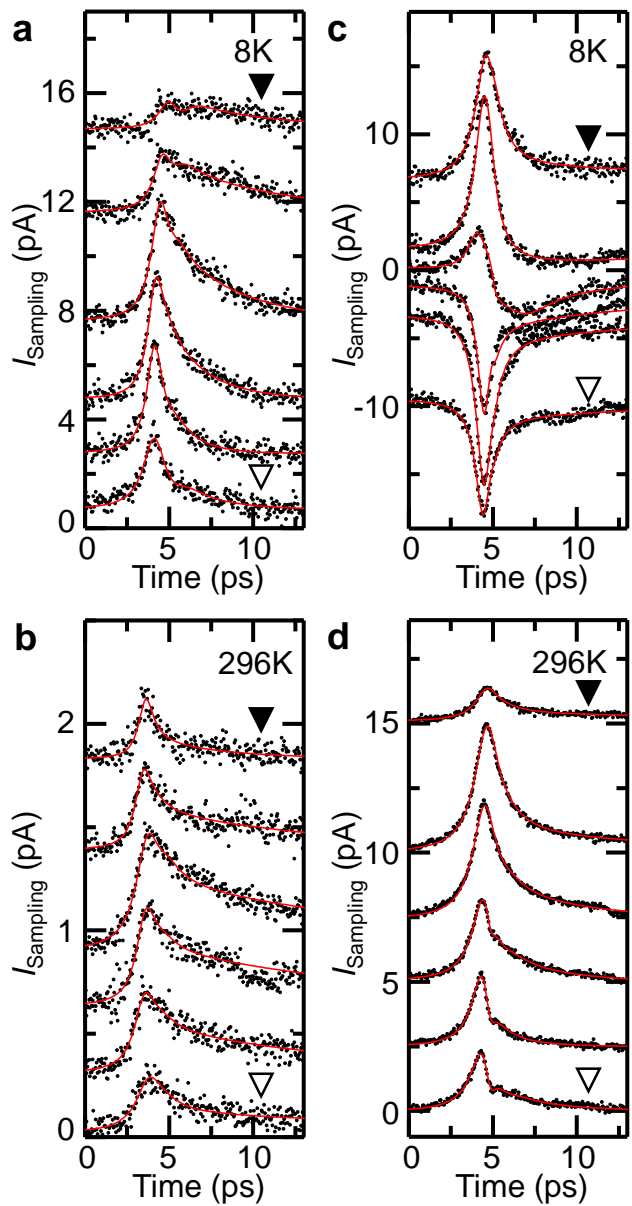

Figure 4

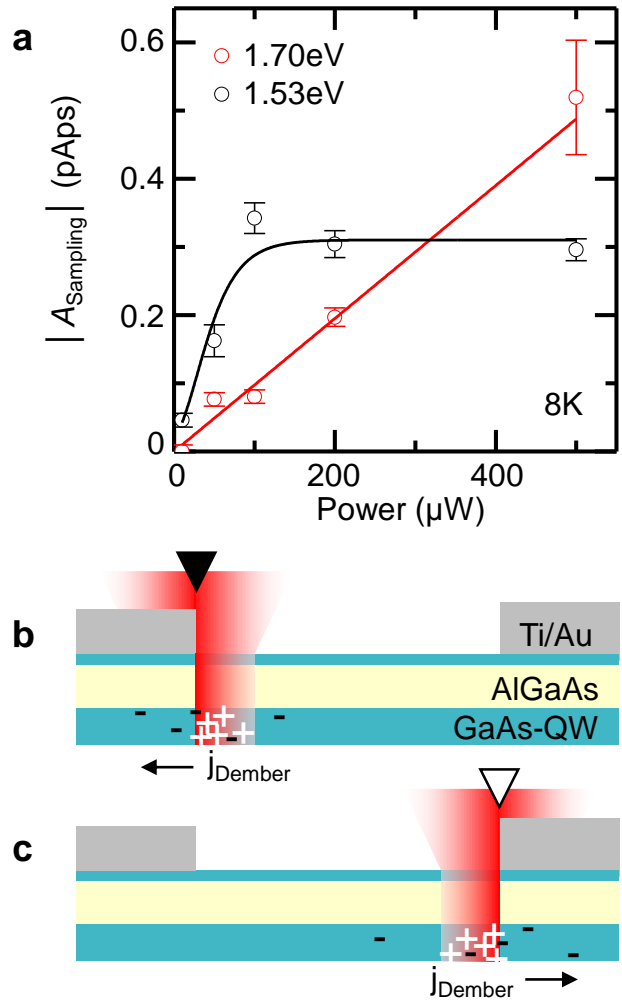

*Figure 5*

# Ultrafast photodetection in the quantum wells of single AlGaAs/GaAs-based nanowires

N. Erhard, S. Zenger, S. Morkötter, D. Rudolph, M. Weiss, H. Krenner, H. Karl, G. Abstreiter, J. Finley, G. Koblmüller, and A. W. Holleitner

- Supplementary Material -

**S1: Nanowire growth parameters.** The GaAs-QW-nanowires are grown in a solid source MBE chamber on a SiO$_2$/Si(111) substrate patterned by nanoimprint lithography (1). The GaAs-cores of the nanowires are grown by catalyst-free vapor-solid (VLS) growth employing a substrate temperature of 610°C, a beam equivalent arsenic pressure BEP(As$_4$) of $1.9 \times 10^{-6}$ mbar, and a Ga flux of 0.25 Å/s for a duration of 150 min. After decreasing the substrate temperature to 490 °C and increasing the BEP to $3.5 \times 10^{-5}$ mbar, the GaAs-cores are overgrown subsequently in situ by Al$_{0.3}$Ga$_{0.7}$As, GaAs, Al$_{0.3}$Ga$_{0.7}$As and a final GaAs capping layer with nominal widths of 30 nm, 2 nm, 30 nm and 10 nm, respectively. The nominal Al content $x$[Al] as well as the nominal thicknesses $d_i$ are obtained from the measured flux ratios and calibrations as described in ref (2).

**S2: Energy dispersive X-ray (EDX) spectroscopy on a GaAs-QW-nanowire.** We perform EDX measurements on the QW-nanowires to confirm the nanowire geometry presented in the main article. Therefore, the nanowires are drop-casted onto C-coated Cu grids and analyzed using a 200kV-JEOL2011 equipped with an EDAX Apollo XL detector. Supplementary Figure S1 a (b) shows a transmission electron microscope (TEM) image of the nanowire tip (base) of a second QW-nanowire grown on the same wafer. The EDX spectra in Supplementary Figure S1 c and d are taken within the area marked with a red circle in Figure S1a and b, respectively. The integration time is set to 100 s. For the nanowire tip, we deduce an Al content of (10 ± 2)% which supports the assumption that there exists an AlGaAs-shell on top of the nanowire. The nanowire base contains (3 ± 2)% Al which is due to the fact that the GaAs-cores are grown first into the pre-defined holes in the SiO$_2$/Si(111) substrate.

**S3 Time-integrated photocurrent measurement.** All presented optoelectronic measurements are performed under vacuum conditions at ~$10^{-6}$ mbar. An objective of a microscope focuses the exciting laser light of a mode-locked titanium:sapphire laser onto the nanowire. The laser spot has a circular shape with a diameter of ~5 μm. The laser is chopped at a frequency of about 2.3 kHz resulting in a photocurrent $I_{\text{Photo}} = I_{\text{on}} - I_{\text{off}}$ detected with a lock-in amplifier utilizing the reference signal provided by the chopper. $I_{\text{on}}$ ($I_{\text{off}}$) is the current through the nanowire device when the laser is switched on (off). (3)

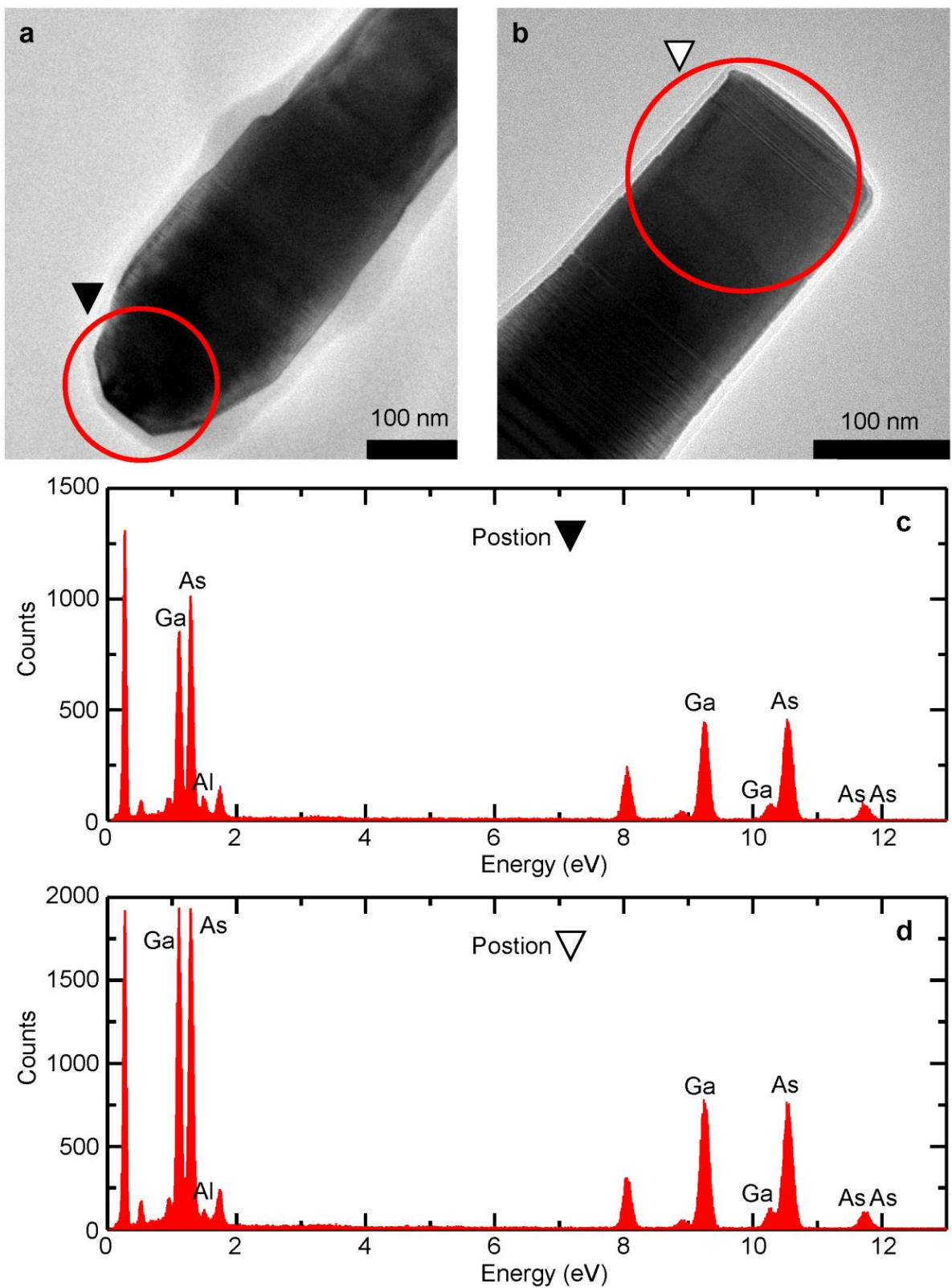

**Supplementary Figure S1:** Transmission electron microscope (TEM) image of the GaAs-QW-nanowire tip (a) and base (b). (c) EDX spectra of the nanowire tip at the position black triangle in (a) and (d) of the nanowire base at the position white triangle in (b).

**S4: Band structure of a contacted GaAs-QW-nanowire.** Supplementary Figure S2 sketches the assumed band structure along a GaAs-QW-nanowire contacted by two gold contacts consistent with the detected photocurrent. The sketch is consistent with all presented measurements. The GaAs-core (green) has a band gap of 1.51 eV at ~8 K (2) and forms an *n*-type Schottky barrier to the gold contact. The AlGaAs-shells (light yellow) have a higher band gap of about 1.92 eV (4) to 1.96 eV (5). For simplicity, we assume that the outer AlGaAs-shell together with the GaAs capping layer forms an *n*-type Schottky contact to the gold which is consistent with the s-shape of the dark current through the nanowire (Figure 2a). However, we note that as we excite only the GaAs in the nanowire, the photocurrent response is not significantly influenced by the contact type of the AlGaAs-shells.

We solve the static Schrödinger equation for the 2 nm GaAs-QW using the effective masses for electrons and holes from ref.(6), and the $Al_{0.3}Ga_{0.7}As$ parameters from ref.(7). Hereby, we obtain the transition energy for excitation from the first heavy (light) hole state to the electron ground state to be ~1.74 eV (~1.77 eV) at 0 K. The value is larger than the 1.68 eV observed in PL- and photocurrent measurements (cf. main article). However, for the calculation, we use bulk values for $Al_{0.3}Ga_{0.7}As$ which may explain the deviation.

The transition energy of 1.68 eV at 8 K is estimated to be ~1.59 eV (~780 nm) at 296 K using the Varshni-equation $E(T)=E(0)-(\alpha T^2/(\beta+T))$ (8). In turn, we use a laser excitation energy of 1.59 eV to resonantly excite the QW at room temperature which is also in accordance with the PL measurements in the main article (Figure 3).

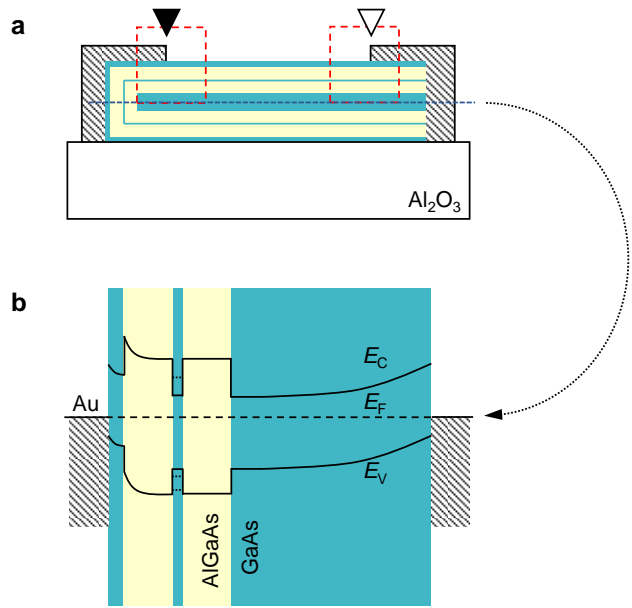

**Supplementary Figure S2:** (a) Schematic drawing of the contacted QW-nanowire lying on a sapphire substrate. The black and white triangles mark the two different contact types as discussed in the paper. GaAs is depicted in green and the AlGaAs in light yellow. The two dashed rectangles highlight the contact regions as sketched in Figure 5 of the main manuscript. (b) Sketch of the band structure of the contacted GaAs-QW-nanowire along the dashed blue line in (a). The bend, dotted arrow on the side panel highlights the corresponding direction within the nanowire. $E_C$ ($E_V$) depicts the conduction (valence) band bottom (top) edge. The dashed line $E_F$ depicts the Fermi level. The dotted lines in the QW mark the first light hole, the first heavy hole and the electron ground state from bottom to top in the QW.

**S5: Thermionic emission and tunneling from the QW into the core through the AlGaAs-barrier.** The emission rate from thermionic tunneling rate $\tau_e^{-1}$ can be estimated as:(9)

$$\frac{1}{\tau_e} = \left(\frac{k_B T}{2\pi m_i d_W^2}\right)^{1/2} exp\left(-\frac{\Delta E}{k_B T}\right),$$

with $k_B$ the Boltzmann constant, $T$ the temperature, $m_i$ the effective mass, $d_W$ the width of the QW, and $\Delta E$ the effective barrier height.

The tunneling rate $\tau_t^{-1}$ from the first sublevel in the QW is given by ref.(10):

$$\frac{1}{\tau_t} = \frac{\pi \hbar}{2 m_i d_W^2} exp\left(-\frac{2 d_B \sqrt{2 m_{bi} \Delta E}}{\hbar}\right),$$

with $\hbar$ the reduced Planck constant, $d_B$ the thickness of the barrier and $m_{bi}$ the effective mass in the barrier.

The calculated transmission times are summarized in the following table. At 300 K thermionic transmission is strong and the photogenerated holes are transmitted to the GaAs-core within the first ps. For 8 K, thermionic emission is strongly suppressed. In addition, the tunneling through the 30 nm thick AlGaAs-shell is weak. In turn, the photogenerated charge carriers stay in the QW within the investigated timescale of 100 ps.

|  | $\tau_e$ at 300 K [s] | $\tau_e$ at 8 K [s] | $\tau_t$ [s] |
|---|---|---|---|
| electron | 4 × 10⁻¹³ | 1 × 10³⁸ | 3 × 10⁻⁴ |
| light hole | 5 × 10⁻¹⁴ | 5 × 10¹⁹ | 3 × 10⁻⁸ |
| heavy hole | 9 × 10⁻¹³ | 4 × 10³⁸ | 2 × 10⁻³ |